\documentclass[11pt,a4paper]{article}
\usepackage[utf8]{inputenc}
\usepackage[T1]{fontenc}
\usepackage[british]{babel}
\usepackage{amsmath, amsfonts, mathtools, amssymb, graphicx, float, lipsum, xcolor, hyperref, authblk, multicol, framed, amsthm}
\usepackage{algorithm, algpseudocode}
\hypersetup{hidelinks}
\usepackage[left=2.00cm, right=2.00cm, top=2.00cm, bottom=2.00cm]{geometry}
\graphicspath{{./figures/}}

\usepackage{array}
\usepackage{booktabs}
\usepackage{cite}
\usepackage{tikz}
\usepackage{float}
\usepackage{tabularx}
\usepackage{adjustbox}
\usepackage{longtable}
\usepackage{rotating}

\newtheorem{example}{Example}[section]

\title{ESSPI: ECDSA/Schnorr Signed Program Input for BitVMX}
\author[1,2]{Sergio Demian Lerner\thanks{\href{mailto:sergio@fairgate.io}{sergio@fairgate.io}}\hspace{3pt}} 
\author[1]{Martin Jonas\thanks{\href{mailto:martin.jonas@fairgate.io}{martin.jonas@fairgate.io}}}
\author[1]{Ariel Futoransky\thanks{\href{mailto:futo@fairgate.io}{futo@fairgate.io}}\hspace{3pt}} 
\affil[1]{Fairgate Labs} 
\affil[2]{Rootstock Labs}
\date{}

\begin{document}
\maketitle

\begin{abstract}
The BitVM and BitVMX protocols have long relied on inefficient one-time signature (OTS) schemes like Lamport and Winternitz to sign program inputs. These schemes exhibit significant storage overheads that hinder their practical application. This paper introduces ESSPI, an optimized method that utilizes ECDSA / Schnorr signatures to sign the input of the BitVMX program. With Schnorr signatures we achieve an optimal 1:1 data expansion, compared to the current known best ratio of 1:200 based on Winternitz signatures. To accomplish this, we introduce 4 innovations to BitVMX: (1) a modification of the BitVMX CPU, adding a challengeable hashing core to it, (2) a new partition-based search to detect fraud during hashing, (3) a new enhanced transaction DAG with added data-carrying transactions with a fraud-verifying smart-contract, and (4) a novel timelock-based method for proving data availability to Bitcoin smart contracts. The enhanced BitVMX protocol enables the verification of uncompressed inputs such as SPV proofs, NiPoPoWs, or longer computation integrity proofs, such as STARKs.
\end{abstract}

\section{Introduction}
Several schemes exist today to optimistically verify Bitcoin computations, such as BitVM2 \cite{1,2}, BitVMX \cite{3}, SNARKnado \cite{4}, and BitSNARK \cite{5}. From these, BitVMX stands out for being the most efficient in terms of onchain cost of dispute, the simplicity of its design, and the generality of the programs it can run without relying on SNARKs. One of the key limitations of BitVMX is that program inputs need to be signed with inefficient OTS schemes such as Lamport or Winternitz. The Winternitz scheme, which is the most efficient of the two, requires on average 25 witness bytes (or vbytes \cite{6}) per signed bit. Unlike standard schemes like ECDSA or Schnorr, which have constant size signatures, Lamport/Winternitz signatures and their public keys grow proportionally with the size of the data to be signed. One of the key open questions related to BitVM protocols is whether there is a more efficient method to sign a BitVM program input. Although the advantage is less significant for BitVM2 that only accepts 300-byte SNARK proofs as input and consumes megabytes in midstates, it is highly relevant to generic verification protocols such as BitVMX that can check not only SNARKs but also larger proofs produced by Nova ($\sim$10 Kbytes) or STARK ($\sim$80 Kbytes). Last, with public-key signed program inputs, BitVMX programs can verify uncompressed inputs, such as BLS signatures for payment channels or NiPoPoWs inputs for bridges. 

The next question is: Can we use a standard signature scheme to sign the program input, benefiting from a constant short signature? Using a standard number-theoretic public-key scheme would not only reduce the on-chain footprint, but also avoid storing and transferring the many one-time public keys required by current BitVM protocols. The benefit is even greater when the BitVMX protocol is extended to $n$ parties, because the number of one-time public keys needed increases quadratically with n. 

In this paper, we positively answer all these previously open questions with a new variant of the BitVMX: ESSPI. The main trick in ESSPI is to use Bitcoin transactions to carry the program input as payload. We sign the hash of the program input twice, once with a public-key scheme and another time with Winternitz, and we prove the equivalence of these two signatures in a second BitVMX instance. This solves the problem of verifying the program input authenticity in the BitVMX CPU. However, because the Winternitz signature only signs the hash of the input, and not every input bit, it creates a new problem: ensuring input data availability. Although embedding signed data in the blockchain can be used to prove data availability and data authenticity to the full nodes, we still need to build a proof mechanism to convince a BitVMX smart contract of data availability. We present another trick to prove the availability of a transaction to a predefined smart contract where the transaction ID is unknown to the smart contract at the time of creation.  

\begin{figure}[h]
\centering
\includegraphics[width=0.8\textwidth]{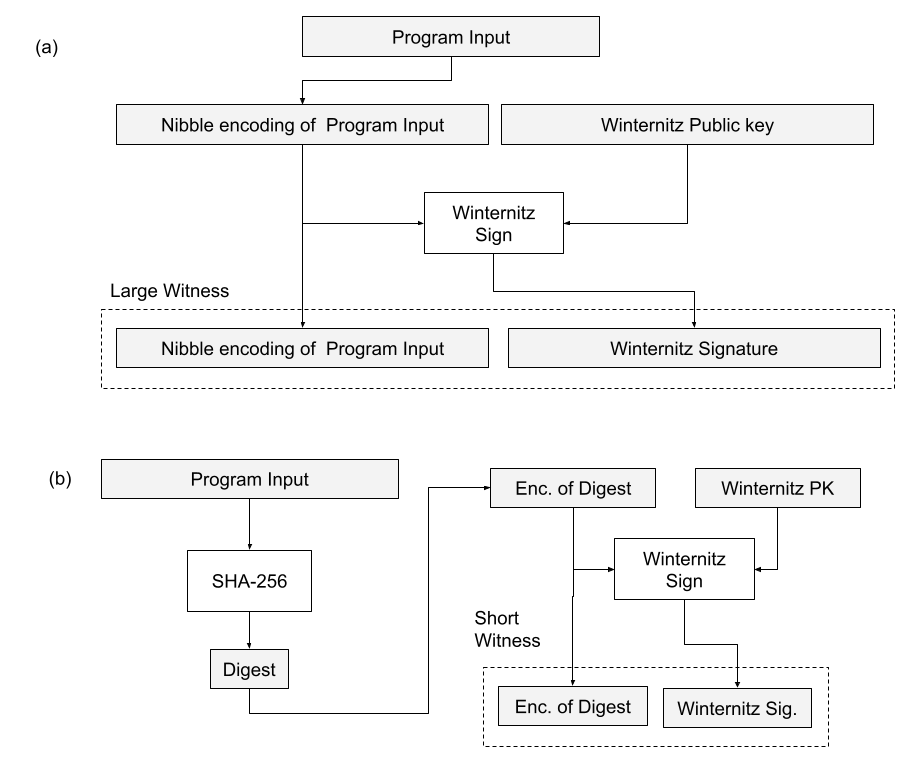}
\caption{The method (a) to authenticate the program input used in the current version of the BitVMX protocol requires long Winternitz public keys and publishes an encoded version of the full program input along the signature resulting in long transaction witnesses. The improved method (b) authenticates the input using a hash digest but does not prove the program input data availability because it doesn't publish the program input inside the witness, only its short digest.}
\label{fig:method}
\end{figure}

The paper is organized as follows. The first section discusses the main problem: how to provide public-key-authenticated data to be used as program input for the BitVMX protocol. The next section lists different ways to use Bitcoin as an efficient data availability layer for large data chunks using standard transactions. In the last section, we show how to build DA proofs specifically for the BitVMX protocol. 

\section{Overview of BitVMX}
BitVMX is an optimistic proving system that allows the verification of arbitrary computations on Bitcoin. Using BitVMX, two parties can create a dispute resolution game on-chain, where they take the respective roles of prover and verifier. The prover can then submit the proof for some arbitrary computation, while the verifier can challenge the proof in case of disagreement. This proof consists of the input and final state of a virtual CPU after executing a pre-agreed program. To ensure that both parties commit to using the BitVMX mechanism, they must both lock some bitcoin as part of the setup process, creating a UTXO whose spending is restricted by the outcome of the protocol. 

The dispute resolution game consists of multiple on-chain rounds where the verifier challenges specific parts of the data that the prover has provided, and the prover must respond with the requested information. The game is designed such that if the proof is invalid, the prover will be forced to commit to conflicting data across different rounds. This results in the prover being unable to respond to at least one challenge, as no valid transaction could be created satisfying the constraints established by the previous transaction scripts. Each round has an associated time lock, and if either party fails to respond within this timeframe, they lose the game, and the other party can claim the funds previously locked. This type of system is called optimistic because the prover wins after a predefined time period if the verifier does not challenge the initial proof, leading to almost no interaction with on-chain transactions. The dispute resolution game is implemented as a directed acyclic graph (DAG) of interconnected transactions, where BitVMX emulates covenants using transactions presigned by all the parties involved. A message-linking scheme based on Winternitz signatures ensures that both parties sign their messages, and relative timelocks are used to force the parties to engage in the game once it is kicked off. Winternitz signatures are also used to authenticate the input of the program.

One of the distinguishing properties of BitVMX is that it verifies the memory consistency using an execution step hash chain instead of using memory Merkelization. When the protocol finds a disagreement step in the execution trace where a memory read is challenged, the prover commits to point of the trace where the memory location was supposedly last written with the read value, and the protocol is able to determine the correctness of this claim by performing another binary search between these two points. 

\section{BitVMX Program Input}
Our goal is to authenticate the input of the BitVMX program (signed by the party that provides it) and prove that it is available to the other parties, so that the verifiers can run the computation locally and be able to challenge a misbehaving prover. The Schnorr and ECDSA signature schemes would be excellent choices to sign the program input because of their short signature size. The seemingly simple task of Schnorr/ECDSA signature verification is difficult in practice due to Bitcoin script limitations. First, the Bitcoin scripting language does not have an opcode to verify that certain data and signature provided as a witness verifies against a predetermined public key. This is what \texttt{CHECKSIGFROMSTACK} \cite{7} does, but it has not been soft-forked into Bitcoin and its future is uncertain.

Second, trying to verify a Schnorr signature with a single program coded using Bitcoin Script is infeasible. One attempt \cite{8} resulted in a $1.1$ GB script, which does not fit into a valid transaction. ECDSA verification requires even more opcodes. 

Third, we cannot use a BitVM protocol to verify the Schorr signature of a message, because we would need to sign the message and the signature with OTS to encode them as program inputs. Clearly, the recursive nature of the attempt can only increase the cost.

A way to bypass these problems is to reuse the Bitcoin transaction signing mechanism to sign the program input. This presents two new challenges. First, the program input payload cannot be signed directly. Instead, the transaction signatures, especially segwit and taptroot signatures, use a hierarchical structure, where the program input can only be inserted in certain spots of this hierarchy, so it gets referenced only indirectly by multiple nested hashes. We can opt to use pre-segwit P2SH addresses, avoiding the hierarchical structure, but we are limited to the storage of data in transaction outputs instead of the witness stack, making program input storage 4 times more expensive in terms of fees, and reducing the storage capacity $4$x. 

The second challenge is even harder to solve: the cheapest method to store information on standard Bitcoin transactions is enveloping, which is the method used by inscriptions. Enveloping involves performing two connected transactions, one to commit to the data and another to reveal it. It is easy to force a party to publish a single data-carrying transaction $D$ by pre-creating a transaction $K$ (Kick-off) having an output that can be consumed by $D$, but also having an alternative time-locked spending path in the same output used to punish that party if it does not publish $D$. To describe such a scheme, we assume that Alice (publisher or prover) and Bob (receiver or verifier) exchange signed transactions containing data elements over the blockchain. We use the following labeling for transactions, signatures, and data elements: an uppercase character indicates the name of the element, the superscript indicates the party that creates and publishes that element ($A$ for Alice and $B$ for Bob) the subscript character indicates what other element it is related to. For example, $P^B_D$ refers to a penalization transaction $P$, issued by Bob, related to the absence of transaction $D$.

Our data publication scheme is depicted in Figure 2. 

\begin{figure}[h]
\centering
\includegraphics[scale=0.4]{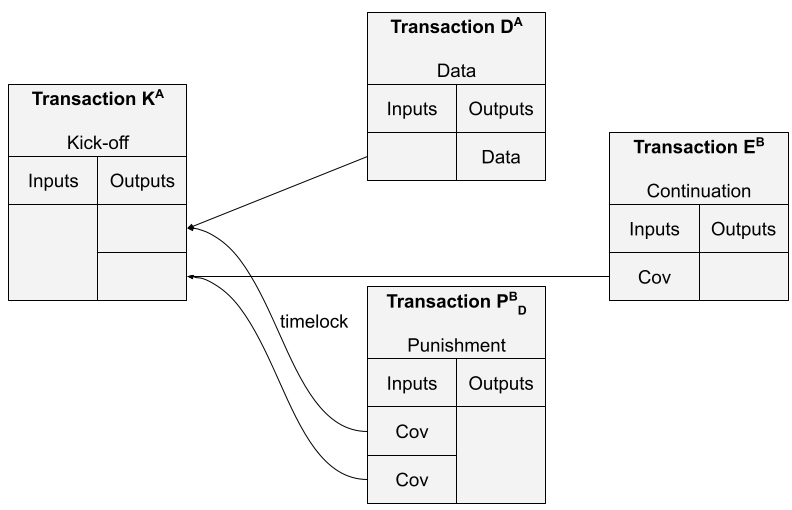}
\caption{Simple Scheme to force publication of Data in Bitcoin}
\label{fig:simple_scheme}
\end{figure}

The alternate path goes into the punishment transaction $P_D$. This transaction also consumes a second output of $K$ that is used to continue or stop the protocol in a transaction $E^B$. The constraint to force the publication of $D^A$ is precreated in $K$. The data is stored in the program outputs of $D^A$. 

Transactions $K^A$, $P^B_D$, and $E$ are pregenerated and cosigned by all parties involved to emulate the covenants. The inputs that require covenant cosignatures are marked with the word \textit{Cov}. Covenants can be implemented with individual signature checks or by a single check derived from a MuSig2 key aggregation. 

Note that this scheme does not work for envelopes. Let us suppose we split the transaction $D^A$ into a commit transaction $C^A$ and a reveal transaction $R^A$. Enforcing the publication of the \texttt{Reveal} transaction $R^A$ (which depends on the commit transaction) is tricky, because the involved parties do not know the commit transaction ID until it is published, so they cannot create the appropriate time-locked spending paths when the protocol is set up. Figure 3 shows a failed attempt to punish the non-revelation of the transaction $R^A$. 

\begin{figure}[h]
\centering
\includegraphics[scale=0.4]{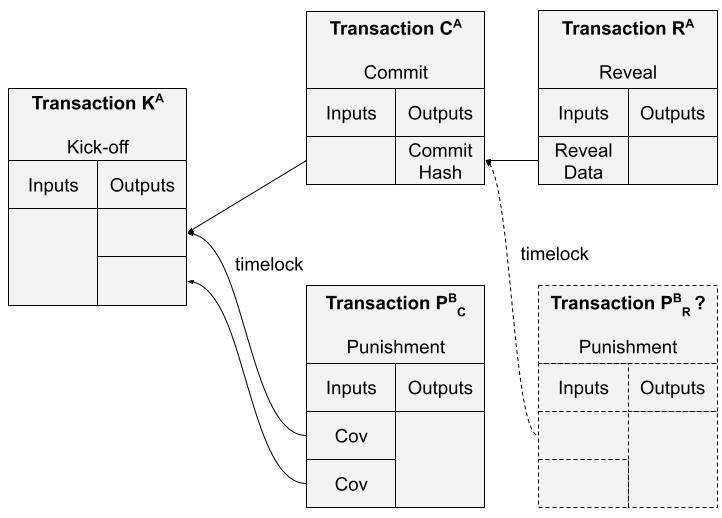}
\caption{A difficult task: The Punishment transaction $P^B_R$ cannot be pre-created because it depends on the transaction ID of $C^A$}
\label{fig:difficult_task}
\end{figure}

Note that $P^B_R$ cannot be pre-created. The opcode \texttt{OP\_CHECKTEMPLATEVERIFY} \cite{9} combined with the opcode \texttt{OP\_CAT} \cite{11} and \texttt{SIGHASH\_ANYPREVOUT} \cite{10} would solve the problem by letting a presigned transaction $PR$ connect to $C$ without knowing its transaction ID. The transaction $P^B_R$ could be restricted to be valid only if it consumes an output from a transaction that, in turn, consumes the first output of $K^A$, which implies that $P^B_R$ is connected to $C^A$. However, these opcodes related to transaction covenants are not available in Bitcoin. 

We now show how to solve all these problems using innovative tools that may also be of interest to other Bitcoin protocols. 

\section{Signed Data Availability on Bitcoin}
The Bitcoin protocol uses the Schnorr/ECDSA signature schemes to sign transactions, but the transaction issuer does not have freedom to store arbitrary data in the transaction, as some fields within a transaction are syntactically or semantically constrained and not intended for arbitrary use. However, Bitcoin provides some flexibility for storing arbitrary data in certain fields of the transaction. One of these fields is the transaction output script. When considering the cost of storing information in a transaction output script, we must take into account the space of the output itself (which is priced at 4 vbytes per data byte) and its dust cost, which is the minimum amount of bitcoins that the output must hold to be considered standard. The dust cost in Bitcoin Core is computed based on the serialized size of the output plus the size of a minimal transaction required to spend it. Currently, this additional size is not computed exactly, but estimated. Bitcoin Core only considers two cases: spending segwit and non-segwit outputs. Segwit outputs pay for an additional $67.75$ data bytes, while non-segwit ones pay for an additional $148$ data bytes ($271$ and $592$ vbytes, respectively). 

We define the expansion factor as the number of vbytes used by transactions per user input byte that needs to be signed, not taking into account constants. The vbyte count comprises space consumed by commitments, public keys, script opcodes for signature verification and data encoding/decoding, and dust costs. The minimum and optimal expansion factor is $1:1$.

We identified many possibilities to store signed data in a transaction:

\begin{enumerate}
    \item \texttt{OP\_RETURN}: Data stored in an output containing an \texttt{OP\_RETURN} opcode in its \texttt{scriptPub}.
    \begin{itemize}
        \item Benefit: Easy to implement and to parse.
        \item Limitation: A standard Bitcoin transaction can only have one \texttt{OP\_RETURN} output and the output can hold a maximum of $80$ bytes. This space is not enough to store a SNARK proof. Storing more data makes the transaction non-standard, but up to $1$ megabyte could be stored in a non-standard transaction. The opcode \texttt{OP\_RETURN} outputs do not consume dust. The approximate expansion factor is $14$x.
    \end{itemize}

    \item \texttt{Enveloping}: Data pushed into the stack in a \texttt{scriptPub} and surrounded by a skipping conditional (\texttt{OP\_PUSH 0} / \texttt{OP\_IF} / \texttt{OP\_ENDIF}). This \texttt{scriptPub} is committed to a P2WSH or P2TR output. P2TR envelopes have the advantage that their scripts are unconstrained, while P2WSH scripts are constrained to $10000$ bytes. The full script is revealed in a following transaction that consumes that output.
    \begin{itemize}
        \item Benefit: a standard transaction can store up to $400$ kilobytes. A non-standard transaction could store up to $4$ megabytes. Also, a transaction can consume multiple outputs where each one reveals $400$ kilobytes of data, and the data is concatenated afterward. This is the method used by Bitcoin inscriptions. The cost of publication is low because the data expansion is close to $1$x.
        \item Limitation: since the script is revealed when the output is spent, we have to use a commit-reveal mechanism similar to the envelopes used for inscriptions. 
    \end{itemize}

    \item \texttt{Annex}: Data in Segwit annex.
    \begin{itemize}
        \item Benefit: In theory, this method could provide unbounded space. The cost could be very low. 
        \item Limitation: The annex is non-standard. It is not forwarded by network nodes.
    \end{itemize}

    \item \texttt{P2WSH Address}: Data stored in multiple standard outputs as (un-owned) addresses. 
    \begin{itemize}
        \item Benefit: Difficult or impossible to censor. 
        \item Limitation: A P2WSH address can store up to $32$ bytes, and an output consumes at least the dust amount. Considering the dust cost the data expansion is approximately $19$x. The benefit is that we can put many P2WSH addresses in a single transaction, so the prevout reference ($160$ vbytes) is amortized over the many outputs. Also a second benefit is that there will be a single signature consuming the handle. 
    \end{itemize}

    \item \texttt{\texttt{scriptPub} with P2PK}: Data can be stored directly in P2PK outputs as $64$-byte public keys. P2PK outputs are considered standard by Bitcoin Core, and no semantic check is made on the content of the public keys while the transactions are in the mempool.
    \begin{itemize}
        \item Benefit: The data expansion factor, considering dust fee, is $\sim 19$x.
        \item Limitations: In the future P2PK outputs may be considered non-standard.
    \end{itemize}

    \item \texttt{\texttt{scriptPub} with bare multisigs}: Bare multisigs using up to $3$ public keys are standard in Bitcoin Core (\texttt{DEFAULT\_PERMIT\_BAREMULTISIG} is true by default).
    \begin{itemize}
        \item Benefit: A bare multisig output can store up to $192$ bytes of data, and considering the dust fee it provides an approximate expansion factor of $13$x.
        \item Limitations: The \texttt{DEFAULT\_PERMIT\_BAREMULTISIG} flag controls bare multisigs forwarding, and can be turned off by nodes.
    \end{itemize}
\end{enumerate}

It is clear that the standardness restrictions make storing information in outputs very expensive.

Note that we do not consider data publication methods where the data is not covered by the ECDSA/Schnorr signature, such as storing data in a pre-segwit \texttt{scriptSig}. Those transactions are malleable and do not provide the authenticity we need. All the methods that store data in output scripts can be complemented with storing an additional byte as the LSB of the output amount, adding an additional random dust cost. 

As mentioned before, it would be superb to have a means to include arbitrary signed data in a transaction without affecting the transaction ID. This is what \texttt{OP\_CHECKSIGFROMSTACK} does, but it hasn't been soft-forked into Bitcoin.

In the following sections, we present two DA methods: the Inclusion-Proof DA and the Timelock-based DA. The Inclusion-Proof DA method proves that a certain transaction was included in the canonical Bitcoin blockchain using cumulative work. Timelock-based DA method forces the prover to sign a transaction before a relative timelock or risks being punished (i.e. loss of BitVMX dispute and security deposit slashed).

Enveloping is the best method to store data for the Inclusion Proof DA, because proving the existence of two transactions (commit and reveal) is almost as expensive as proving the existence of one transaction. It is also best for Timelock-based DA, although it is not possible with the tools defined so far. We will present the Timelock-based DA using transaction outputs to store the program input to simplify the explanation, as it requires a single Schnorr signature to be verified. Although using signatures to store program input data is more efficient, the use of multiple Schnorr signatures forces the need to verify a SNARK inside BitVMX to prove cheating. Finally, we will show how to use enveloping for Timelock-based DA, which is optimal.

To summarize, given that the public key of the Schnorr signed data can be provided to all the parties involved in the BitVMX protocol in advance, a Schnorr signed transaction can be used to broadcast the input to all participants in a way that:

\begin{enumerate}
\item All participants receive the program input data.
\item The input data is signed with one or more predefined public keys known to all participants. 
\end{enumerate}

\begin{example}
If a $100$ kilobyte STARK proof is split into standard \texttt{OP\_RETURN} outputs, it requires $1250$ standard transactions each having a single \texttt{OP\_RETURN} output, it also requires $1250$ output handles to attach the transactions, and $1250$ signatures. If the STARK proof is split into P2WSH addresses, it requires a single transaction with $3125$ outputs. Since a standard transaction cannot accommodate $3125$ outputs, but only $2500$, it would need to be split into $2$ transactions, consuming two handles with two different signatures. The maximum storage capacity for a single transaction using P2WSH outputs is $80$ kilobytes. If the STARK proof is stored using transaction envelopes, then the whole proof can fit in \texttt{scriptPub} stored in a single transaction.    
\end{example}

Until now we have proved that the program input data is authentic and available to full nodes but we need to prove it to a smart contract running in Bitcoin, either by creating a non-interactive proof that can be checked by BitVMX or by other means. In other words, we only need a way to use the signed data as input of the BitVMX protocol such that using the wrong data during RISC-V computations can be challenged as easily as Winternitz signed data, and this is achieved by a variation of the BitVMX protocol we now present.

\section{Proving Data Availability to BitVMX}
We've guaranteed authenticity and availability of the program input to full nodes. Now we want to prove these two facts to a BitVMX protocol instance. We can prove a transaction has been mined and confirmed by many blocks using an SPV proof, and we rely on the difficulty and cost of mining a hidden fork. We can also prove it by punishing the prover if they do not publish the program input before the BitVMX dispute begins using a relative timelock. In this case, we rely on the difficulty and cost of transaction censorship. These two proofs seem to be based on different crypto-economic assumptions. However, to allow SPV proof to protect high value, an autonomous system that relies on SPV proofs must accept and compare proofs with counter-proofs. Since the time to provide counter-proofs must be bounded, the security of this method also relies on the assumption of non-censorship of transactions. Since the whole BitVMX protocol also relies on eventual non-censorship, it is acceptable to use simply the DA method based on timelocks. 

Nevertheless, we present two methods to prove data availability: Inclusion-Proof DA (based on the cost of mining) and Timelock-based DA (based on the cost of censorship)

\subsection{Inclusion-Proof DA}
The protocol needs to make sure that the verifier has received the input so that it can challenge it. The standard method is that a certain transaction was included in the blockchain and that it has been confirmed by a number of headers (or cumulative work). This is often called a SPV proof. An instance of BitVMX can be used to verify the SPV proof, which could be stateful or stateless, allowing counter-proofs or not \cite{12}. If the SPV proof is incorrect, Bob can challenge it. We will not go deeper into the many variants of SPV proofs and will present the simplest possible protocol for comparisons.

To support a SPV proof for DA, we use a two-stage protocol:

\begin{enumerate}
    \item A first BitVMX instance proves input data availability by verifying the SPV proof. If data availability is not challenged for a certain time, then the prover has the opportunity to continue with a second BitVMX instance which actually uses the data proven to exist. The SPV proof convinces the verifier of the existence of a transaction in the canonical Bitcoin chain. The proof can be created using a NiPoPoW, a SNARK, a STARK, Nova, or any other argument, but generally it is compressed with a SNARK because of the high cost of signing program inputs with an OTS. Assuming the input of the DA proving instance is a SNARK, and this SNARK is signed by Winternitz OTS, the input consumes approximately 60K vbytes.
    \item The second BitVMX instance receives as input a hash of the data proved to be available in the first instance. This requires signing $32$ bytes with the Winternitz OTS, which consumes only $6.4$ K vbytes.
\end{enumerate}

The SNARK proof of data availability will contain (as hidden witness) the Merkle Inclusion Proof for the transaction containing the program input, its block header, and a number of headers confirming the first one adding up a certain amount of pre-established cumulative work.

Although variants of this method are used by all Bitcoin rollup bridges based on BitVM2 that have been designed so far (BitVM2 bridge, AlpenLabs' Strata bridge, and Citrea's Clementine), it is complex and requires proof/counter-proof interactivity \cite{13} to reach the desired level of security to protect high bitcoin amounts. 

\subsection{Timelock-based DA}
We present a novel method to prove data availability and sign program input for BitVM-like protocols. We start with a simple protocol that uses ECDSA signatures, and later we present a protocol that uses Schnorr signatures. As before, two parties, Alice (prover) and Bob (verifier), exchange signed transactions containing data elements using the Bitcoin blockchain. 

We define User Input (UI) as the input the user program will need to consume to decide the outcome of the BitVMX protocol (accept or reject the spending). For example, if the user program must check a SNARK, then the UI is exactly the SNARK proof. The Program Input (PI) will be a message that can be accessed by the BitVMX CPU and contains the UI, but may also contain additional padding, header, or footer that should be skipped by the user program. In other words, the program must parse the PI to extract the UI.

A first kick-off transaction K contains a predefined P2SH output called \textit{handle} that contains two spending paths (using \texttt{OP\_IF}/\texttt{OP\_ELSE}/\texttt{OP\_ENDIF}). The first path is used by a transaction DA that provides the User Input (UI Data). 

The second path is used by a cosigned penalization transaction $P^B_D$, and it has a relative timelock. Consuming the handle with transaction $D^A$ before the timelock requires a signature $S$ with a private key owned by Alice. This key should be used to sign a single instance of the transaction DA and must not be reused.  Figure $4$ depicts the DAG of transactions that are used.

\begin{figure}[h]
\centering
\includegraphics[width=0.8\textwidth]{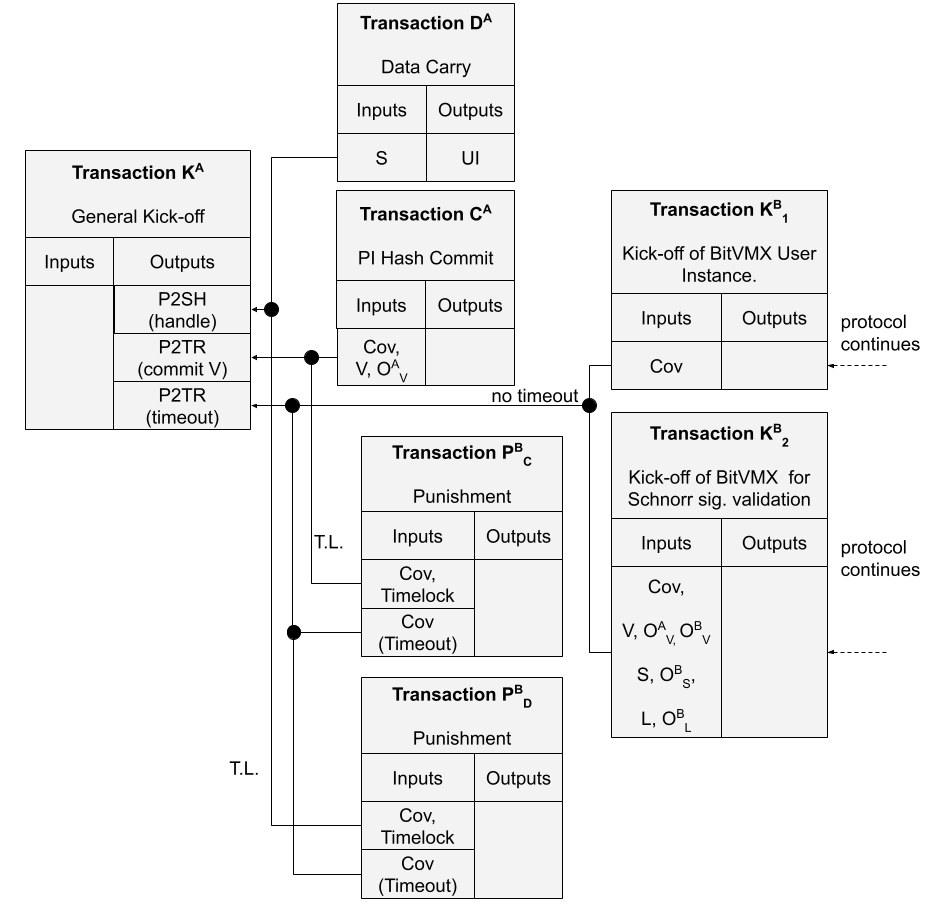}
\caption{The part of the transaction DAG of a BitVMX kick-off that accepts ECDSA-signed UI embedded in the Program Input}
\label{fig:transaction_dag}
\end{figure}

Transactions $K^A$, $P^B_C$, $C^A$, $P^B_D$, $K^B_1$, and $K^B_2$ are presigned by both participants emulating covenants. Each cosigned input is indicated in the diagram with the word \textit{Cov}. $P^B_C$ and $P^B_D$ are penalization transactions if Alice does not publish $C^A$ or $D^A$, (respectively) before a relative timelock $T$ is enabled. Once the kick-off transaction $K^A$ is published, Bob can punish Alice if Alice does not consume the handle after a time $T$. The BitVMX protocol kick-off transaction $K_1$ may be delayed a maximum of $T$ time. Since the handle is only constrained by an \texttt{OP\_CHECKSIGVERIFY}, Alice can consume the handle in any way she wants to avoid being punished by the timelock, but other checks will restrict the transaction $D^A$. Alice must also publish another transaction $C^A$ which commits to the hash of the Program Input ($V$) using an OT signature $O^A_V$. The OTS signature $O^A_V$ is verified against a OT public key established by transaction $K^A$. 

The same protocol can be implemented using multiple handles to support multiple data-carrying transactions, each one providing a part of the program input, but to simplify our presentation, we restrict ourselves to using a single handle.

Although the BitVMX program could somehow check that the signature is correct as a prelude, we instead use fraud proofs. A secondary instance of the BitVMX protocol ensures that the signature of the program input matches the value $V$ committed and signed with the OTS signature $O^A_V$. If the transaction is malformed, Alice will lose the protocol challenge in the second BitVMX instance. For example, the transaction $D^A$ could contain one \texttt{OP\_RETURN} with $80$-bytes of payload, multiple outputs with bare multisigs to store more data, or a non-standard $1$ megabyte \texttt{OP\_RETURN} payload. 

We now formally specify the validations performed by the second BitVMX instance. We define the message $D'$ as the signed message for legacy inputs. It is derived from $D^A$ using the Bitcoin legacy P2SH signing rules. The rules remove script data from all inputs other than the one signed and insert  \texttt{scriptPub} into the input where the signature was. The BitVMX Program Input will be exactly $D'$. It is the responsibility of the program running on BitVMX to extract the User Input (UI) from $D'$. Figure 5 shows how the Program Input is memory-mapped for a transaction carrying an \texttt{OP\_RETURN} with the User Input. 

\begin{figure}[h]
\centering
\includegraphics[width=0.8\textwidth]{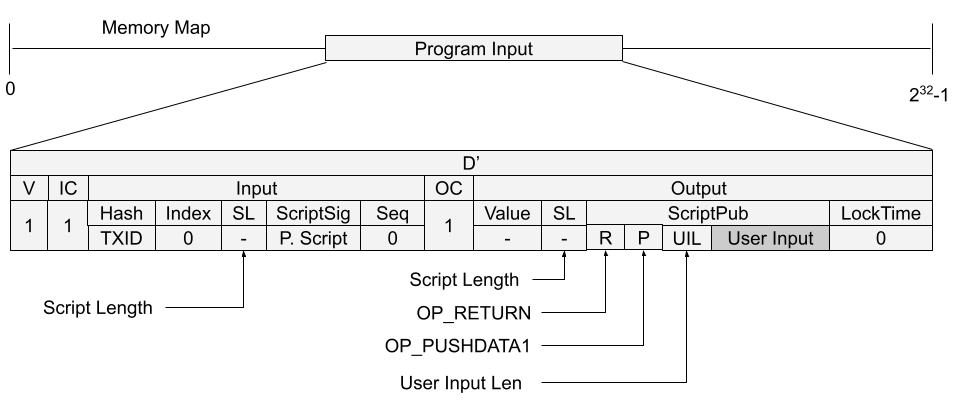}
\caption{The User Input is embedded in the Program Input, which is memory mapped by a 32-bit BitVMX CPU.}
\label{fig:user_input_embedded}
\end{figure}

This instance of BitVMX must be set up to accept up to $1$ megabytes of ECDSA-signed program input, as this is the maximum size of $D'$ according to the Bitcoin consensus rules. 

The value $V$ must match the hash digest L which is the single SHA-256 hash of the message $D'$. If we apply a second SHA-256 hash to $V$, we obtain the "signature hash", which is the value actually signed by the ECDSA signature $S$. The signature $S$ is the one provided as \texttt{scriptSig} in the only input of the transaction $D^A$. 

Figure 6 shows how the value $V$ is computed from $D^A$. Alice computes and OT-signs $V$.

\begin{figure}[h]
\centering
\includegraphics[width=0.8\textwidth]{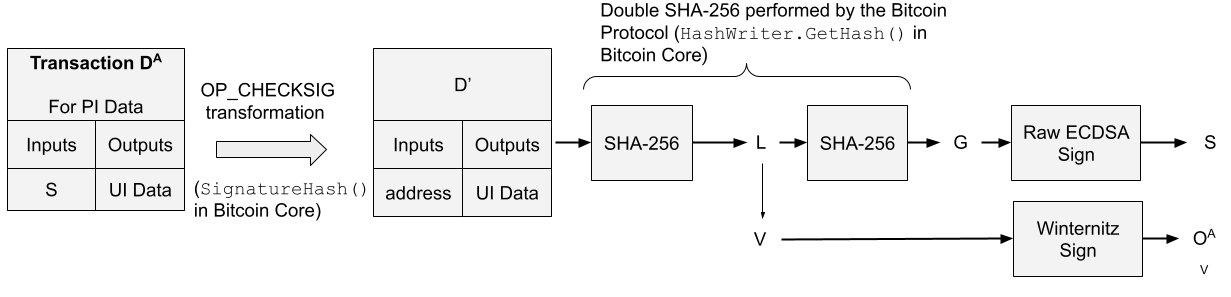}
\caption{The figure depicts the different transformations that the transaction $D^A$ undergoes when being signed by ECDSA by the Bitcoin Protocol, and how Alice extracts $V$ for OT-signing.}
\label{fig:transaction_transformations}
\end{figure}

The raw ECDSA sign operation represents the ECDSA signature of a $32$ byte value without pre-hashing, so the diagram shows all hash operations involved. The value $L$ represents the middle state inside the Bitcoin double-hash operation, while the verifier computes $L$ and will verify that it matches the signature $S$.

If Bob detects that $V$ provided by Alice doesn't match $L$ (the single SHA-256 hash of the program input $D'$), he launches an instance of BitVMX where he submits as program input the values $V$ (OT signed by both parties with $O^A_V$ and $O^B_V$), $S$ (alleged raw signature of $G$, SHA-256 of $V$, provided by Alice) signed by Bob ($O^B_S$) and $L$ (single SHA-256 of $D'$ calculated by Bob) also signed by Bob ($O^B_L$). Bob wins this protocol if he can provide those three values $V$, $S$, $L$ such that:

\begin{enumerate}
\item $V$ is correctly OT-signed by Alice with $O^A_V$, and $V$, $S$ and $L$ are correctly OT-signed by Bob with $O^B_V$, $S^B_V$ and $L^B_V$. If this does not hold, the transaction providing the program inputs will be invalid, as OT signatures are verified by Bitcoin script. 
\item $S$ is a valid ECDSA signature of $G$, the hash of $L$, where the signature is verified against Alice's public key. Here, the program performs a single SHA-256 hash of $L$ to get $S$ to check the raw signature. If this validation does not hold, then Bob loses the challenge.
\item $L$ is not equal to $V$. If this is true, then Bob wins, as he could prove fraud. 
\end{enumerate}

Figure 7 shows the checks performed by the secondary BitVMX instance. Note that Check 1 is performed by the Bitcoin script, not the BitVMX program.

\begin{figure}[h]
\centering
\includegraphics[scale=0.4]{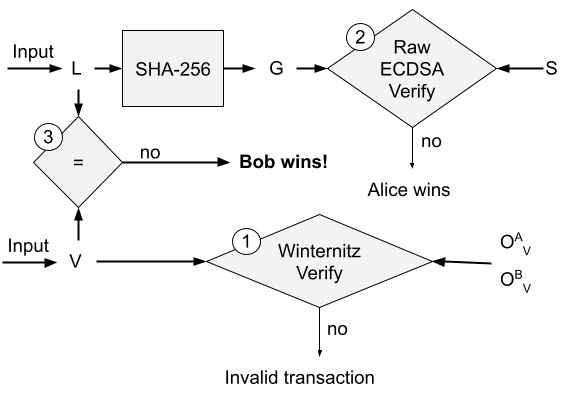}
\caption{The $3$ ordered checks performed by the secondary BitVMX instance.}
\label{fig:secondary_bitvm_checks}
\end{figure}

In this new instance of BitVMX, the roles on this new BitVMX are inverted, and it is Alice that must challenge Bob to prevent Bob from winning. We don't need to execute SHA–256, just ECDSA verification and a simple equality check.

It is possible to reuse the same BitVMX instance (DAG and OT public keys) for the user-program and the ECDSA signature validation if Bob signs a program input bit that forces the instance to switch program code, but for simplicity, we will present the protocol using two BitVMX instances set up for the two different purposes.

Figure $4$ shows how the two existing BitVMX instances (primary and secondary) are connected. After Alice publishes a general kick-off transaction $K$, Bob can either continue with the primary BitVMX protocol (transaction $K^B_1$), launch a secondary BitVMX protocol to challenge the signature (transaction $K^B_2$), or punish Alice for not publishing the input (transaction $P^B_C$).

In this first protocol, we use a legacy address for the handle and get $D'$ as the program input with the user input embedded directly. Because our BitVMX design can only check sequentially hashed program inputs against an OT signed digest $V$, what the Program Input and $V$ represents depends on how the User Input is actually hashed. Depending on what type of address is used for the handle, we get different program input types. The value $V$ will always be the single SHA-256 hash of the PI. Table $1$ shows some of the possibilities, excluding transaction envelopes.

\begin{table}[h]
\centering
\begin{tabular}{|l|l|l|}
\hline
\textbf{Address type of the handle} & \textbf{UI storage} & \textbf{Program Input Type} \\
\hline
P2TR & Script & Tapleaf tagged message \\
\hline
P2TR & Tx output & a \texttt{CTxOut} structure \\
 & & (referenced by \texttt{sha\_single\_output}) \\
\hline
P2WSH & WitnessScript & a \texttt{scriptPub} \\
\hline
P2WSH & Tx output & a \texttt{CTxOut} structure \\
 & & (referenced by \texttt{hashOutputs}) \\
\hline
P2SH & \texttt{scriptSig} & impossible because the \texttt{scriptSig} is not signed \\
\hline
P2SH & Tx output & a modified transaction \\
\hline
\end{tabular}
\caption{The table shows what the Program Input represents depending on the type of address used for the handle and where the UI is stored within the data transaction.}
\label{tab:program_input_types}
\end{table}

Like the Lightning Network and many of the BitVM protocol variants, the protocol may be the target of transaction replacement cycling attacks \cite{14}, as the handle can be spent by both Alice and Bob after the time-lock. Alice can try to avoid being punished by submitting transactions with increasing fees and switching inputs, Since the value $V$ is committed by the transaction $C^A$, any replacement transaction $D_i$ that does not match the hash $V$ will make Alice lose the dispute. Therefore, Alice can only use CPFP and change the spent/unspent state of another input to grind the transaction. To solve this, a special function is added to check the correct format of the transaction $D'$ to the program that runs in the first instance, and this function is executed first. The function must check that $D'$ only has one input and the required data-carrying outputs.

\section{BitVMX for OT-Signed Program Input Hash}
Our approach to reduce the program input size relies on OT-signing a hash digest of the program input data instead of all the program input data. If the hash digest represented the root of a Merkle tree whose leaves include all the program input data words, then this tree could be directly mapped to a standard Merklelized RAM\cite{15}\cite{16}, as commonly used in many memory-consistency systems for optimistic CPUs such as Cartesi\cite{17}. In this case, we can map the program input into a specific part of the RAM, fill the remaining memory with zeros, and build a tree that represents the initial state of the RAM. No change in the optimistic CPU is required. However, the Bitcoin transaction does not Merklelize its script data, and the signature covers a sequential hash of the script data. The taptree does contain Merklelized script data, but only one leaf is made available, so it is not useful for our purposes.

If Alice OT-signs a RAM Merkle root for the CPU to start with, Bob needs to be able to challenge Alice in case the OT-signed RAM Merkle root does not contain the program input published in the DA transaction. Bob can use the secondary BitVMX instance to prove fraud by providing a SNARK or he can challenge Alice to provide a SNARK that proves that no fraud occurred (depending on which party we prefer to be paying the price of the proof). If Bob does not challenge the root hash, it means that both parties have agreed that the program input represented by the hash root is correct, and the optimistic CPU executes as normal. This solution is conceptually simple, but still requires using a SNARK, which adds more complexity to the system, and may require additional cryptographic and trust assumptions.

The BitVMX CPU uses a new memory consistency system that minimizes the number of hash operations required to verify a memory access, but BitVMX does not Merkelize the memory \cite{15}. This means that either we change BitVMX to support Merklelized memory or we must find a method to upgrade the BitVMX CPU to verify sequential hashing.  We opt for the latter option. To this end, we present a new CPU mode for BitVMX that enables reading from unsigned program inputs.  To verify the correctness of the unsigned program input,  it must be sequentially hashed resulting in the hash digest $V$, and we assume that both parties have agreed on the correctness of $V$ (the Schnorr/ECDSA signature $S$ has not been challenged). 

\subsection{The ICM CPU Mode}
We define a special BitVMX CPU mode called Input Check Mode (ICM) to support the Schnorr-signed program input. While the CPU is in ICM mode, the executed program will scan the unsigned input, hashes it, and checks the hash digest against a given hash $V$.

To specify the ICM mode, we begin with the specification of the BitVMX protocol, as presented in its white paper.

In the BitVMX white paper, the execution trace is defined as:
\[
\mathtt{trace}_i = \mathtt{write.address}_i \| \mathtt{write.value}_i \| \mathtt{writePC.pc}_i
\]

The full trace is defined as:
\[
\mathtt{full trace}_i = \mathtt{read1.address}_i \| \mathtt{read1.value}_i \| \mathtt{read1.lastStep}_i \|\cdots\| \mathtt{writePC.pc}_i
\]

The step hash is defined is:
\[
\mathtt{stepHash}_i = h(\mathtt{stepHash}_{i-1} \| \mathtt{trace}_i)
\]

We define two new buffers in RAM. The Midstate Buffer (MIB) contains 32 bytes, and it represents either the midstate or last state of the SHA-256 compressor. When the CPU starts, it contains the first starting state of SHA-256. The MIB is memory-mapped. 

The Message Buffer (MEB) is a $64$ byte buffer that is also memory-mapped, and the code can freely read or write this part of the memory.

We change the trace to include the MIB. 

We add two new opcodes to the CPU: \texttt{HASH\_UPDATE} and \texttt{HASH\_FINAL}. The opcode \texttt{HASH\_UPDATE} uses the SHA-256 one-way compression function (OWCF) to hash the MEB into the MIB, continuing from the prior state of the MIB.

To simplify the hashing core, it is the responsibility of the program code to add the appropriate padding to the message in the MEB before finalizing the hash with \texttt{HASH\_FINAL}, including in the last block the correct message bit count. Note that the bit count can depend on the unsigned program input as long as the unsigned data is read and parsed after it has been re-written with \texttt{LSSW}, so that there is a write operation prior to any read of the UPI. 

The protocol is simplified if the PI byte count is signed in the SPI. In this case, we can enforce that the PI has a length that is multiple of $64$ bytes, and the padding is straightforward. For the bit count to be automatically computed, it would need to be tracked by the CPU and stored in some part of the Midstate Buffer instead of being provided in the last MEB block by the program. 

If there can only be a short amount of trailing data after the UI in the PI, then it is also possible to avoid the computation of the padding in the BitVMX program.

To use this alternative method, we force the UI length to be a multiple of $64$ bytes, truncate the hash computation to the last midstate prior the padding block (which will not contain UI data), and check $V$ against a midstate instead of to a final SHA-256 hash. The trailing data must be short because to prove the matching with the script or transaction hash in the secondary BitVMX instance, one party would need to submit the trailing data, possibly signed with OT signatures. 

In the ICM the CPU can use four memory regions: 

\begin{enumerate}
\item UPI (Unsigned Program Input): Holds the unsigned program input and this data is RAM memory-mapped
\item SPI (Signed Program Input). This is a normal OT signed program input (by Alice). When using UPI, this area will be used to store a single hash digest. It is also RAM memory-mapped.
\item MEB (Message Buffer). This is a small $64$ byte buffer that is used to store the message input to the SHA-256 function.
\item MIB (Midstate Buffer). This $32$-byte buffer stores the midstate or final state of the SHA-256 function.
\end{enumerate}

There are two kinds of program input: unsigned program input (UPI) and signed program input (SPI). Both are read-only memory-mapped sections of RAM. For clarity, in this paper we exclude program inputs OT-signed by the challenger or Schnorr/ECDSA-signed by the challenger, as the protocol can be easily extended to support them without affecting the protocol soundness. 

During ICM the program will only access the MEB, UPI and the CPU registers. There is no need to enforce access boundaries by the CPU because the prelude code is agreed on by Alice and Bob, and both can check its correctness. 

The number of instructions executed in ICM mode can be fixed (i.e., step $1$ million) or specified by Alice and signed with OTS. The simplest method is to reserve a number of steps for the ICM. The last instruction of the ICM should be a \texttt{HASH\_FINAL}.

To make effective use of the ICM, the parties agree on a prelude program called the ICM program that linearly hashes the UPI and leaves the hash digest in the Midstate Buffer. However, to hash the UPI, the ICM program uses the embedded instructions \texttt{HASH\_UPDATE} and \texttt{HASH\_FINAL} and not a SHA-256 subroutine coded in RISC-V. 

Note that we can let normal execution (non-ICM) use the internal hasher embedded in the CPU by adding an additional opcode \texttt{HASH\_RESET}. The program would be able to reset, append, and finalize the hasher at any time. Also, in non-ICM mode, we do not restrict the position of the hash update instructions. Adding a hasher to the CPU reduces the size of any program that makes heavy use of hashing, such as a STARK or Bitcoin SPV proof verifier.

\subsection{Execution Trace Sections}
Both Alice and Bob must agree on the execution of a program having two sections $A$ and $B$. The section $A$ is executed in ICM, while the section $B$ is executed normally.

\begin{itemize}
    \item Section $A$: The CPU starts in ICM mode. In this section the program uses the Message Buffer, the Midstate Buffer and the UPI, as previously defined. During this section, the program will read the UPI, move the bytes to the MEB, perform hashing operations using \texttt{HASH\_UPDATE} finalize hashing with \texttt{HASH\_FINAL}, and leave the final result in the MIB. For each word read from UPI, the program copies it into the MEB and then writes it back into the same UPI address. We have to make sure that the UPI read and MEB write instructions are back-to-back to be able to challenge both instructions simultaneously in case the hash chain fails at the read. To simplify our presentation, we will introduce a single machine instruction to do both. We define a new machine instruction \texttt{LSSW} (for load/store/store word) that reads a word from UPI memory, writes it back to the same address and simultaneously writes it to the MEB, with the MEB offset being the address modulo $64$ (the size of the MEB). It will not store the value in any destination register. 
    
    \begin{example}
        For example, on a RISC-V BitVMX CPU a normal program that tries to read the UPI would be read using the \texttt{lw rd}, \texttt{imm(rs1)} instruction, store the result in register \texttt{rd}, and then move it to the MEB with a \texttt{sw} instruction. Instead, we will use the new instruction \texttt{LSSW imm(rs1)}. For this instruction the BitVMX protocol will generate a trace with the format: 
        \begin{align*}
            \mathtt{read1.address}_i &== \mathtt{imm(rs1)}\\
            \mathtt{write.address}_i &== \mathtt{read1.address_i}\\
            \mathtt{read1.value}_i &== \text{word stored in UPI at offset } \mathtt{imm(rs1)}\\
            \mathtt{write.value}_i  &== \mathtt{read1.value_i}
        \end{align*}

    If the word read is not fully contained in the UPI memory area, the instruction will halt the CPU.
    When challenging a read/write operation, the \texttt{LSSW} instruction will be also valid as if the following write trace had been produced:
    \[
    \mathtt{write.address}_i == (\mathtt{read1.address}_i - \mathtt{UPI\_base\_Offset}) \% 64 + \mathtt{MEB\_base\_offset}
    \]
    No word from the UPI will be read with any other instruction.
    \end{example}
    
    \item Section $B$: This is where the CPU goes back to normal. This section can verify a SNARK or execute any other user-defined program.
\end{itemize}

As mentioned above, the easiest way to separate each section is to establish a fixed boundary of each section. If Alice will not use all the steps available in a certain section, she can fill the unused steps with NOPs.

The point $AB$ (boundary between section $A$ and $B$) contains the latest state of the state buffer and the Midstate Buffer, which should match the hash $V$.

\subsection{Disagreement Search Protocol}
First, we note that the verifier (Bob) knows $V$, because it is signed by OTS in the dispute kick-off transaction $K^A$. Once Bob has access to $V$, he may realize that $V$ does not match the program input signed by Schnorr/ECDSA in the Bitcoin transaction $D^A$. On mismatch, Bob will launch an instance of BitVMX to prove fraud and punish the prover (Alice). If Alice has lied about $V$, she will lose the dispute in this secondary BitVMX instance. Assuming Bob has not disputed $V$, the parties can continue assuming that $V$ is valid and sufficient proof that the signature $S$ exists. Therefore, there is no need to actually verify $S$ in the BitVMX primary instance. 

We expand the trace so that it also includes the MIB and the opcode. 
\[
\mathtt{trace}_i = \mathtt{write.address_i \| write.value_i \| writePC.pc_i \| MIB_i \| opcode_i}
\]

Note that if the CPU performs many writes per instruction, this structure plus SHA-256 padding may not fit into a single $64$ byte message block, and so the protocol will need to support challenging larger message hashes, which is possible in Bitcoin Script. Otherwise, we can strip the SHA-256 padding block, since we are hashing a message with known size. For a $32$-bit RISC-V CPU, the trace structure requires $49$ bytes, so it can fit into a single SHA-256 block nicely with standard padding.

First, we assume that Bob disagrees with the last step hash of execution, and he wants to prove fraud during the execution. Bob knows that there must be a step he agrees with followed by another she disagrees with, between step 0 and the last step.

\subsection{Partition Search}
Bob starts partitioning the step hash chain until he finds the first point $r$ where she agrees on step $r-1$ but disagrees on step $r$. If $r > AB$, then it means that all the computations performed during Section $A$ are correct. If the read memory address in step $r$ is in the UPI, then the value must have been written during section $A$ or $B$, since the chain is correctly computed up to step $r-1$, the read value must be correct. If Alice shows an incorrect trace value for the step where the memory word is written, then the hash chain will be challenged as normal in BitVMX (an invalid hash chain slot cannot produce a valid slot later without breaking the hash function).

Now we must analyze the case where $r < AB$. This means that Alice is trying to write or read incorrect values to the UPI during section $A$. Because Alice has signed $V$, we know that the state of the MIB at step $AB$ must be $V$, if not, then Bob can prove fraud by showing $O^A_V$ to a script. 

Bob will request the traces for steps $r-1$ and $r$, and he gets the opcode Alice claims is executed at step $r$. If the trace for step $r-1$ does not match the step hash, he will challenge the hash operation, as normal in BitVMX. Therefore, we now assume $MIB_{r-1}$ is correct. 

Then, if the instruction is neither \texttt{LSSW} nor \texttt{HASH\_UPDATE} it will be challenged with the normal instruction challenges of BitVMX. If the midstate changes but the opcode is not a hashing one, then Bob can also challenge it.

If the instruction is a \texttt{HASH\_UPDATE} then we take $x = r$ and skip the part where Alice pushishes $x$.

If the instruction at step $r$ is not \texttt{HASH\_UPDATE} then it must be \texttt{LSSW} and we execute the following sub-protocol. Bob will ask Alice to give the $x$ (after $r$) which corresponds to the next \texttt{HASH\_UPDATE} or \texttt{HASH\_FINAL} opcode, Alice will also respond with a trace for the $x$ step (including $MIB_x$), $MEB_x$, $x$ itself, and the $MIB_x$ signed by Alice. If the OWCF of $MEB_x$ from $MIB_{r-1}$ does not yield $MIB_x$, Bob will challenge the OWCF and the script will validate the hash. Now Bob is assured that $MIB_x$ is a result of the OWCF that continues the midstate $MIB_{r-1}$. Alternatively, the hash operation could be challenged by Alice only if the input/output values signed by Alice do not satisfy the OWCF function.  

Note that Bob may detect that the next hash update in her trace is not in step $x$, but he will play along with Alice's response anyway.

If $MEB_x$ is correct, then Bob will challenge the value written by \texttt{LSSW} to the MEB. Because the program code will write each word exactly once between hash updates, the written value must be wrong. Let the offset in MEB written by \texttt{LSSW} be \texttt{ofs}, then Bob will provide $r$, $MEB_x[\texttt{ofs}]$, $x$ (all co-signed by Alice) to prove that $\mathtt{write.value_r} \neq MEB_x[\texttt{ofs}]$. 

Now we assume that $MEB_x$ is incorrect. In this case $MIB_x$ cannot be incorrect, because a previous script verified that it was computed hashing $MEB_x$, which now we assume to be incorrect. For $MIB_x$ to be correct, then Alice would need to have found a second preimage of the hash compression function. 

Now we assume to have an incorrect $MIB_x$ and $MIB_x$. $MIB_x$ cannot correspond to a future or past valid midstate of the chain, because the script proved that it used $MIB_{r-1}$ as the starting state, and we can assume that Alice cannot find cycles on the hash chain. Therefore, $MIB_x$ must correspond to a state unknown to Bob and not any past or future midstate. Then Bob will challenge the continuation of the hash. Bob knows that $MIB_{AB}$ is correct, because this must match the value $v$ signed by Alice. If $MIB_x$ is incorrect, then Bob can perform a binary search on the traces between step $x$ and step $AB$, until he finds two steps $z, z + 1$ where $MIB_z$ is incorrect but $MIB_{z+1}$ is correct. 

Now Bob will challenge Alice to provide any $64$ byte MEB value that can be applied to state $MIB_z$ to produce state $MIB_{z+1}$, which Alice would not be able to do without breaking the second-preimage resistance property of the hash function. Figure $8$ shows a flowchart with the decisions and actions Bob takes to win the dispute.

\begin{figure}[h]
\centering
\includegraphics[scale=0.5]{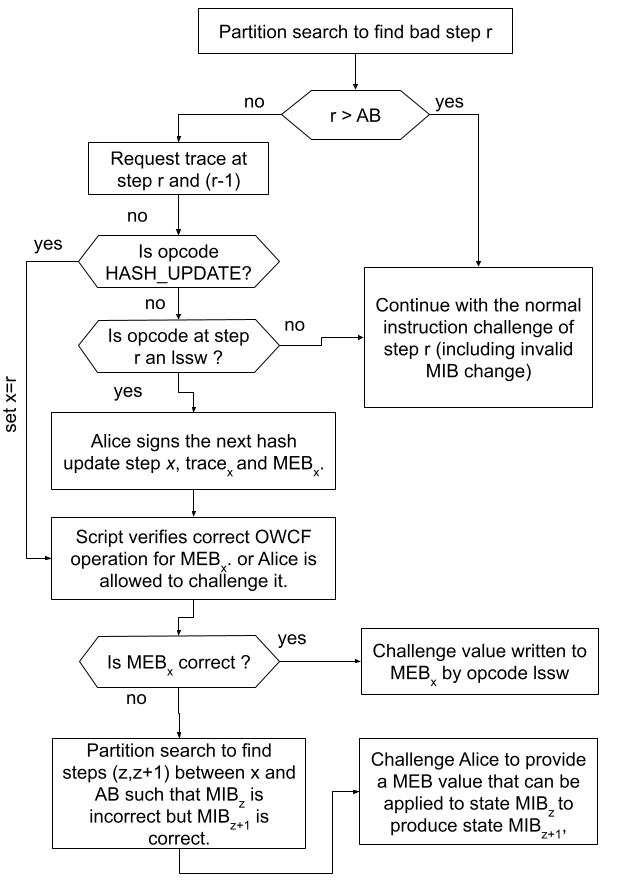}
\caption{The flowchart with decisions and actions for Bob to win the dispute on the sequential hashing of the program input.}
\label{fig:dispute_flowchart}
\end{figure}

\section{Using Enveloping for the Timelock-based DA Scheme}
We now describe a two-party BitVMX that allows envelopes to be used to store the program input. A multi-party version of this protocol requires additional tools. The protocol starts with a kick-off transaction $K$ that has two outputs: the handle and the timeout signal. The handle is used to publish the program input. The timeout output is used to penalize Alice and prevent the protocol from continuing. This is a simplification of the BitVMX protocol, as penalization in BitVMX may also involve burning or seizing Alice's security bond. We also leave out of this paper the details on how the locked bitcoins are given to Alice if she wins the dispute.

The data is published using transaction envelopes with the transactions $C^A$ (commit) and $R^A$ (reveal). The transactions $K^B_1$ and $K^B_2$ are the kick-offs of two instances of BitVMX (primary and secondary). The primary is used for the normal user-defined program, the secondary is used by Bob to prove fraud in how the transactions $C^A$ and $R^A$ were built. The transactions $K^A$, $K^B_1$, $K^B_2$, and $P^B_C$ are pre-created and they are co-signed with Schnorr when published. The transactions $C^A$, $R^A$ and $P^B_D$ are created only when the program input needs to be committed.

\subsection{The DA-DAG}
We first make $V$ (the script hash signed by Bob) the tapleaf-tagged hash associated with the script $U$. Set $V = \mathtt{tagged\_hash}(``TapLeaf", bytes([leaf\_version]) + \mathtt{ser\_script}(U))$. According to BIP-$340$, the message actually hashed by SHA-256 will be $M$. Set
\[
M = \langle \mathtt{tag\_hash + tag\_hash + leaf\_version + compact\_size}(U) + U \rangle
\]
The field \texttt{tag\_hash} is the SHA-256 of the string \texttt{TapLeaf}. Figure 9 shows how the User Input is embedded in the Program Input, which is memory mapped.

\begin{figure}[h]
\centering
\includegraphics[width=0.8\textwidth]{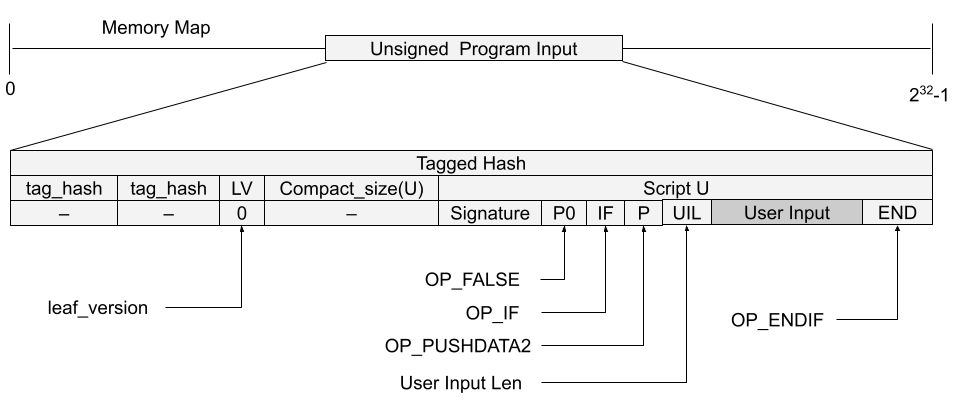}
\caption{When using enveloping, the User Input is embedded in the tagged hash, which represents the Program Input for a 32-bit BitVMX CPU.}
\label{fig:user_input_embedded_in_tagged_hash}
\end{figure}

This message $M$ will be the Program Input mapped into the CPU memory. The BitVMX program will see the actual User Input prefixed with a header that ends with an \texttt{OP\_PUSHn} instruction and the User Input length. The program must decode the UI from $M$ to use it. Note that P2WSH and P2TR addresses are built by hashing the script with a single SHA-256 application, and not two, which is the standard pre-segwit. 

We now present the transaction DAG to enable the use of transaction envelopes. The P2TR address $X$ specifies a Script Merkle Tree. The taproot internal key for $X$ should be an unspendable \textit{Nothing Up My Sleeve} (NUMS) point, i.e., a point with unknown discrete logarithm. The taptree of $X$ must contain two leaf nodes for two scripts, one checking the signature $W$ against a known public key $PK^W$, and having a relative timelock, and another committing to a script $U$ that contains the user's input to the program. This leaf stores the tagged hash $V$. 

If the P2TR address $X$ is replaced by a P2WSH address, the verification is simplified as $V$ would be directly the hash of witnessScript and not a tagged hash, but the script size would be constrained to $10000$ vbytes by the Bitcoin consensus rules. 

Figure 10 shows the DAG of transactions required to use enveloping as a mechanism to publish the User Input (UI) inside the Program Input.

\begin{figure}[h]
\centering
\includegraphics[width=0.8\textwidth]{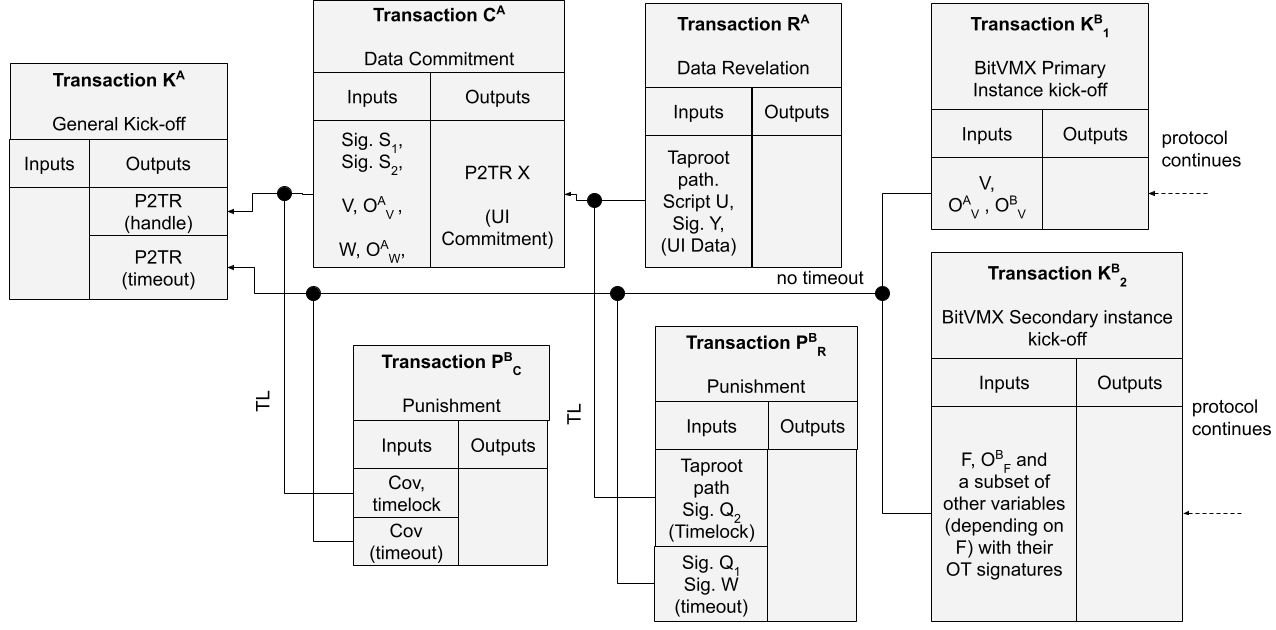}
\caption{The transaction DAG for using enveloping to publish the BitVMX Program Input.}
\label{fig:transaction_dag_for_enveloping}
\end{figure}

In our previous scheme, we introduced a secondary BitVMX instance ready for Bob to prove that the signature S was invalid. Now we use a similar secondary BitVMX instance to challenge different kinds of fraud attempts, not only about the signature, but also about the transaction format. For example, we must allow Bob to prove that Alice built $C^A$ in a way that prevents Bob from penalizing Alice if she does not publish or $R^A$. Since neither $C^A$ nor $R^A$ are pre-created, Alice has too many degrees of freedom on how to build them, but Bob must enforce they are built to such that he can penalize Alice if:

\begin{enumerate}
\item Alice doesn't publish $R^A$
\item Alice builds $R^A$ in a way that prevents Bob from obtaining a lineal hash of the PI from a specific place in $C^A$.
\item The signature $S$ is malformed.
\end{enumerate}

From now on, we slightly modify the definition of the $D'$ notation. We use transaction letters with apostrophes to refer to the associated fixed-size data structures (the signed message) that are actually signed by Schnorr in taproot addresses.

The value $W$ is a signature of the transaction $P^B_R$ associated with a public key owned by Alice. $W$ is used in the input that spends the \texttt{continue/stop} signal. Let $Q_1$ be a signature of the transaction $P^B_R$ that is associated with a public key owned by the verifier for the \texttt{continue/stop} signal, and $Q_2$ be a signature of the transaction $P^B_R$ by the verifier for the output of $C^A$.  

All public keys are known at the setup time. Note that the first input of $P^B_R$ does not require a signature from Alice: If the timelock expires, Bob is free to spend it without stopping the protocol, but this can only go against his interests. In a multiparty ($n > 2$) setting all verifier's signatures would be required.

In the notation used below $O_W$ means the OTS signature of $W$, while $sig.W$ means that the Schnorr signature $W$ of the current transaction is verified. 

The transaction $C$ receives as witness $W$, and $OW$. $W$ is the transaction ID of $C^A$, and $O^A_W$ is the OTS of $W$. It is perfectly possible for a transaction to provide its own transaction ID as a witness because the transaction ID does not hash the elements of the witness stack.  

The transaction $C^A$ is signed twice with Schnorr, resulting in the signatures $S_1$ and $S_2$, both stored in the witness. 

We define $C'$ as the taproot signed message, which is a message derived from $C$ that is finally signed. It is hashed with SHA-256 once, and then the digest is signed by the raw Schnorr algorithm. The signed message varies depending on the transaction input and the sighash flag. For taproot inputs, it is composed of a tagged hash message with $\mathtt{tag} =\mathtt{TapSigHash}$ and
\[
\mathtt{data} = \mathtt{sighash epoch} + \mathtt{common signature message + extension}
\] 
and the common signature message is derived from $C$. Table 2 shows the composition of the common signature message with the script path spend extension.

\begin{table}[H]
\centering
\small
\begin{adjustbox}{max width=\textwidth}
\begin{tabularx}{\textwidth}{|l|l|X|}
\hline
\textbf{Field} & \textbf{Size} & \textbf{Description} \\
\hline
\texttt{hash\_type} & 1 & A byte indicating the which inputs/outputs are being signed \\
\hline
\texttt{nVersion} & 4 & The transaction version field. \\
\hline
\texttt{nLockTime} & 4 & The transaction locktime field. \\
\hline
\texttt{sha\_prevouts} & 32 & The SHA-256 hash of the txid+vout outpoints for all the inputs included in the transaction. $\sim$\texttt{SIGHASH\_ANYONECANPAY} \\
\hline
\texttt{sha\_amounts} & 32 & The SHA-256 hash of all the output amount fields for all the inputs included in the transaction. $\sim$\texttt{SIGHASH\_ANYONECANPAY} \\
\hline
\texttt{sha\_scriptPubkeys} & 32 & The SHA-256 hash all the output \texttt{scriptPubkeys} for all the inputs included in the transaction. $\sim$\texttt{SIGHASH\_ANYONECANPAY} \\
\hline
\texttt{sha\_sequences} & 32 & The SHA-256 hash of all the sequence fields for all the inputs included in the transaction. \\
\hline
\texttt{sha\_outputs} & 32 & The SHA-256 hash of all the outputs in the transaction. $\sim$(\texttt{SIGHASH\_NONE} or \texttt{SIGHASH\_SINGLE}) \\
\hline
\texttt{spend\_type} & 1 & A single byte that encodes the extension flag and annex present values. \\
\hline
\texttt{outpoint} (input) & 36 & The \texttt{txid} + \texttt{vout} outpoint of the input being signed for \texttt{SIGHASH\_ANYONECANPAY} \\
\hline
\texttt{amount} (input) & 8 & The amount field of the input being signed for \texttt{SIGHASH\_ANYONECANPAY} \\
\hline
\texttt{scriptPubKey} (input) & variable & The \texttt{scriptPubkey} of the input being signed for \texttt{SIGHASH\_ANYONECANPAY} \\
\hline
\texttt{nSequence} (input) & 4 & The sequence field of the input being signed for \texttt{SIGHASH\_ANYONECANPAY} \\
\hline
\texttt{input\_index} & 4 & The vin of the input being signed for. $\sim$\texttt{SIGHASH\_ANYONECANPAY} \\
\hline
\texttt{sha\_annex} & 32 & The SHA-256 of the optional annex included at the end of the witness field. \\
\hline
\texttt{sha\_single\_output} & 32 & The SHA-256 of the output opposite the input currently being signed for \texttt{SIGHASH\_SINGLE} \\
\hline
\texttt{tapleaf\_hash} & 32 & The leaf hash for the chosen script you're using from the script tree. Script path spend extension (\texttt{tapscript}) \\
\hline
\texttt{key\_version} & 1 & The type of public key used in the leaf script. Script path spend extension (\texttt{tapscript}) \\
\hline
\texttt{codesep\_pos} & 4 & The opcode position of the last \texttt{OP\_CODESEPARATOR} in the leaf script. Script path spend extension (\texttt{tapscript}) \\
\hline
\end{tabularx}
\end{adjustbox}
\caption{The common signature message with the Script path spend extension.}
\label{tab:common_signature_message}
\end{table}

Due to the additional complexity of the DA-DAG compared with our previous scheme, we describe it formally. We make use of an auxiliary function to create Taproot addresses and create Winternitz signature verification scripts:

\begin{enumerate}
    \item \texttt{TaprootAddress}(list-of-scripts): a P2TR address with a tree containing the given scripts with unspendable NUMS internal key.
    \item \texttt{\texttt{OT-CSIGV}}($M$, list of OT-public keys): a script that reads from the stack an encoded input message $M$ and one or more OT signatures and verifies them against a list of OT public keys. Aborts the script if any of the verifications fail. 
\end{enumerate}

We define a transaction as $\langle \text{list-of-inputs, list-of-outputs} \rangle$ and each input as $\langle \text{prevout, s} \rangle$ where $s$ is the script leaf used by this input from the ones committed in prevout. Each script $s$ should appear twice in the DAG, one when committed and another when referenced. We omit versions, amounts, and nSequence values and taproot paths for simplicity. We additionally use:

\begin{itemize}
\item "-" to indicate that we don't care what inputs/outputs are present in the transaction,
\item $T.o[i]$ to indicate the output with index $i$ of $T$ starting from $1$.
\item Covenant-Check to indicate the checking of two signatures of the transaction, one of each party either separately or combined with MuSig2.
\item \texttt{CSIGV}$(x)$ is defined as for the script that pushes value $x$, followed by \texttt{CHECKSIGVERIFY}.
\item \texttt{CSEQV}$(x)$ is defined as the script that pushes value $x$,  and executes  \texttt{CHECKSEQUENCEVERIFY} followed by \texttt{DROP}.
\item The \texttt{sver\_} prefix for scripts that verify that a certain variable has been pushed and signed correctly.
\end{itemize}

$OPK^A_W$, $OPK^B_W$, $OPK^A_V$, $OPK^B_V$, $OPK^B_{S1}$, $OPK^B_{S2}$, $OPK^B_F$, $OPK^B_X$, $OPK^B_L$ are the one-time public keys related to signatures $O^A_W$, $O^B_W$, $O^A_V$, $O^B_V$, $O^B_{S1}$, $O^B_{S2}$, $O^B_F$, $O^B_X$, $O^B_L$, respectively. The superscript indicates the party who owns the public key.

$PK^A_W$, $PK^A_Y$, $PK^A_{S1}$, $PK^A_{S2}$, $PK^B_{Q1}$, $PK^B_{Q2}$ are the Schnorr public keys related to the signatures $W$, $Y$, $S_1$, $S_2$, $Q_1$, $Q_2$.

The DA-DAG is defined as
\begin{align*}
\texttt{script\_}K^A_{O1}C^A &= \langle\texttt{CSIGV}(PK^A_{S1}), \texttt{CSIGV}(PK^A_{S2}), \texttt{\texttt{OT-CSIGV}}(OPK^A_V) , \texttt{\texttt{OT-CSIGV}}(OPK^A_W) \rangle\\
\texttt{script\_}K^A_{O1}P^B_C &= \langle \texttt{CSEQV}(T) ,  \text{Covenant-Check} \rangle\\
\texttt{sver\_}V &= \langle \text{\texttt{OT-CSIGV}}(V,  \langle OPK^A_V , OPK^B_V \rangle) \rangle\\
\texttt{sver\_}F &= \langle \text{\texttt{OT-CSIGV}}(F, OPK^B_F ) \rangle\\
\texttt{sver\_}E_C &= \langle \text{\texttt{OT-CSIGV}}(EC , OPK^B_{EC} ) \rangle\\
\texttt{sver\_}W &= \langle \text{\texttt{OT-CSIGV}}(W,  \langle OPK^A_W , OPK^B_W \rangle) \rangle\\
\texttt{sver\_}S1 &= \langle \text{\texttt{OT-CSIGV}}( S1, OPK^B_{S1} \rangle ) \rangle\\
\texttt{sver\_}S2 &= \langle \text{\texttt{OT-CSIGV}}( S2, OPK^B_{S2} \rangle ) \rangle\\
\texttt{sver\_}C' &= \langle \text{\texttt{OT-CSIGV}}( C', OPK^B_{C'} \rangle ) \rangle\\
\texttt{sver\_}C^A &= \langle \text{\texttt{OT-CSIGV}}( C, OPK^B_C \rangle ) \rangle\\
\texttt{sver\_}X  &= \langle \text{\texttt{OT-CSIGV}}( X, OPK^B_X \rangle ) \rangle\\
\texttt{sver\_}L  &= \langle \text{\texttt{OT-CSIGV}}( L, OPK^B_L \rangle ) \rangle\\
\texttt{sver\_}R^{A'} &= \langle \text{\texttt{OT-CSIGV}}( R^{A'}, OPK^B_{RA'} \rangle ) \rangle\\
\texttt{sver\_}E_R &= \langle \text{\texttt{OT-CSIGV}}(ER , OPK^B_{ER} ) \rangle\\
\texttt{script\_}F_{(1)} &=  \langle \texttt{sver\_}F, \texttt{sver\_}E_C,  \texttt{sver\_}W , \texttt{sver\_}S_1, \texttt{sver\_}S_2 \rangle \\
\texttt{script\_}F_{(2)} &=  \langle \texttt{sver\_}F, \texttt{sver\_}C', \texttt{sver\_}V \rangle \\
\texttt{script\_}F_{(3)} &=  \langle \texttt{sver\_}F, \texttt{sver\_}W, , \texttt{sver\_}S_1, \texttt{sver\_}C^A \rangle \\
\texttt{script\_}F_{(4)} &=  \langle \texttt{sver\_}F, \texttt{sver\_}R^{A'}, \texttt{sver\_}Y, \texttt{sver\_}E_R \rangle \\
\texttt{script\_}K^A_{O2}K^B_1 &= \langle \texttt{sver\_}V \rangle \\
\texttt{script\_}K^A_{O2}P^B_C &= \langle  \text{Covenant-Check} \rangle \\
\texttt{script\_}K^A_{O2}P^B_R &=  \langle \texttt{CSIGV}(PK^B_{Q1}), \texttt{CSIGV}(PK^B_W) \rangle \\
\end{align*}

\begin{align*}
K^A &= \langle - ,\langle \texttt{TaprootAddress}(\langle \texttt{script\_}K^A_{O1}C, \texttt{script\_}K^A_{O1}P^B_C \rangle), \\&
\texttt{TaprootAddress}( \texttt{script\_}K^A_{O2}P^B_C , \texttt{script\_}K^A_{O2}P^B_R , \texttt{script\_}K^A_{O2}K^B_1 , \texttt{script\_}F_{(i)} \\&
\text{ for } (1\le i \le4)) \rangle \rangle\\
\texttt{script\_}C^A_{O1}P^B_R &= \langle \texttt{CSEQV}(T) , \texttt{CSIGV}(PK^A_{Q2}) \rangle\\
\texttt{script\_}U &= \langle \texttt{CSIGV}(PK^A_Y), \texttt{PUSH 0}, \texttt{IF}, \texttt{PUSH  [user input]}, \texttt{ENDIF} \rangle\\
C^A &= \langle \langle \langle K^A.o[1] , \texttt{script\_}K^A_{O1}C^A \rangle \rangle,  \langle \texttt{TaprootAddress}(\langle \texttt{script\_}U , \texttt{script\_}C^A_{O1}P^B_R) \rangle \rangle\\
R^A &= \langle \langle \langle C^A.o[1] , \texttt{script\_}U \rangle \rangle, \langle\rangle \rangle \\
K^B_1 &= \langle \langle \langle K^A.o[2] , \texttt{script\_}K^A_{O2}K^B_1 \rangle \rangle, - \rangle\\
K^B_2 &= \langle \langle \langle K^A.o[2] , \texttt{script\_}F(i) \text{ for any } (1\le i \le4)\rangle \rangle , - \rangle \\
P^B_C &= \langle \langle \langle K^A.o[1] , \texttt{script\_}K^A_{O1}P^B_C  \rangle, \langle K^A.o[2] , \texttt{script\_}K^A_{O2}P^B_C \rangle \rangle, \langle\rangle \rangle \\
P^B_R &= \langle \langle \langle C^A.o[1] , \texttt{script\_}C^A_{O1}P^B_R  \rangle, \langle K^A.o[2] , \texttt{script\_}K^A_{O2}P^B_R  \rangle \rangle, \langle\rangle \rangle \\
\end{align*}

In the Kick-off transaction $K^B_2$, Bob must select the type of fraud he is trying to prove by specifying a fraud index ($F$), which he OT-signs with $O^B_F$. Each challenge requires consuming a different tapleaf from the input taproot tree, and each leaf script accepts and checks a specific $F$ value and the OT signatures of the arguments required by that fraud proof. We now list the challenges that must be supported to cover all types of fraud that Alice could attempt. Each assumes that the previous challenge was not required (that is, the challenge number $2$ assumes that the transaction $C$ is well formed).

\subsubsection{Invalid Transaction $C^A$}
\begin{itemize}
    \item Program inputs: $E_C$, $V$, $W$, $S_1$ and $S_2$. Note that Bob already has $O^A_W$, and $O^A_V$ so it only needs to co-sign them. Here $E_C$ represents the actual fields of $C$ that are required to build $C'$ and validate the signatures $S_1$ and $S_2$ (amounts, pubkeys, prevouts, etc.). 
    \item Goal: Bob wants to prove that transaction $C^A$ is correctly signed by $S_1$ or $S_2$, but it is malformed and does not contain a valid P2TR address $X$, or it contains additional unexpected inputs or outputs. This is the hardest fraud to prove because we need to prove $C^A$ is malformed without giving a literal description of $C^A$. 
    \item Checks: If the signature $S$ was created with \texttt{SIGHASH\_ALL}, the field \texttt{sha\_outputs} ($32$) included in $E_C$ is the SHA-256 of the serialization of all transaction outputs, each one in \texttt{CTxOut} format (amount,\texttt{scriptPub}Key). Since the number of outputs is not committed in $C'$, we need a method to prove that the transaction $C^A$ has an invalid number of outputs without listing all outputs. We want the program input to be bounded in size and small. We provide three of such methods. 
    The first method is to prove that the last hashed message block added to the SHA-256 function while processing \texttt{sha\_outputs} contains a bitcount that doesn't match the expected value. This can be done by extending a freely-chosen SHA-256 midstate, updating the midstate with the last padding block and finalizing the hash function so that the result matches the known hash digest. A match in hash digests proves that the bitcount value present in the padding block is actually correct and not part of the user message itself. However, this technique requires the additional cryptographic assumption that SHA-256 is secure from free-start collisions. 

    The second is based on the fact that during setup, the transaction ID of $K^A$ is known to both parties, also are the input and output amounts of $C^A$ and the handle address. The ID can be embedded into the secondary program code or ROM, or it can be co-signed with the OTS. Finally, the values $W$ and $V$ are signed by Alice, which means that the program in the secondary BitVMX instance that receives $W$ and $V$ can rebuild both the single input and single output of $C^A$, hash each one, and compute both the correct \texttt{sha\_outputs} and \texttt{sha\_prevouts} fields. Then it can compare those against the ones in $C'$. A mismatch in \texttt{sha\_outputs} would indicate that the output of $K^A$ has an incorrect address, or there are additional unexpected outputs. A mismatch in \texttt{sha\_prevouts} would indicate there are additional inputs, or Alice has reused the public key $PK^A_S$ in another unrelated transaction signature. The only downside is that the setup phase gets more complex due to the interdependences. 

    The third method, the one we have specified in this paper, is based on $K^A$ requesting not a single signature $S$ from Bob, but two signatures $S_1$ and $S_2$, related to different public keys $PK_{S_1}$ and $PK_{S_2}$, in the same transaction input. The signature $S_2$ must sign using the \texttt{SIGHASH\_SINGLE} flag, which only stores in \texttt{sha\_single\_output} the \texttt{CTxOut} hash of a single output that contains $X$. The signature $S_1$ must sign without this flag, which stores in \texttt{sha\_outputs} the hash of all outputs. If \texttt{sha\_outputs} is equal to \texttt{sha\_single\_output} then this implies the transaction has a single output. If an incorrect sighash flag is used for any of the two signatures, then Bob can easily prove this by showing $C'$, without listing the actual outputs, because the \texttt{hash\_type} field indicates this. 

    Once the single \texttt{CTxOut} hash has been identified, the fraud checking program must check that the address $X$ is related to a taptree (the taproot scripts Merkle tree) that contains exactly two leaves with scripts built from $W$ and $V$ that are given as arguments. The leaf related to $W$ must commit to a script containing a relative timelock and require a single Schnorr signature $W$ related to $PK_W$ (the public key used by $W$). The other leaf related to $V$ must match the tapleaf hash $V$ ($V$ already hashes the tag, script size and script hash). The leaves must be sorted according to the numerical order of their tagged hashes. The output amount must also be checked against a pre-established hardcoded value, because the transaction $P^B_R$ must consume this exact amount. It can also be allowed to vary if Alice publishes and signs with a OTS the amount. The program must also check that the internal key of $X$ is a predefined unspendable NUMS.

    The fields present in the Script path spend extension (\texttt{tapleaf\_hash}, \texttt{key\_version}, \texttt{codesep\_pos}) do not need to be checked because they are already constrained by the scripts in the handle output of $K^A$. 

    The unexpected inclusion of the \texttt{sha\_annex} can be detected and proven in by exhibiting the annex hash and wrong \texttt{spend\_type}. 

    Potentially invalid \texttt{nLockTime} and \texttt{nSequence} values or input amounts can also be checked directly in this challenge, but invalid values can't harm Bob, as Bob obtaining $C^A$ implies the transaction has already been mined. 

    Finally Bob needs to prevent the inclusion of a large number of inputs in $C^A$ without listing those prevouts. We solve this problem in a manner similar to the outputs, by requiring in $K$ two signatures for $C^A$, $S_1$ and $S_2$ (the same can be reused). The signature $S_1$ must sign all prevouts \texttt{SIGHASH\_ALL} and $S_2$ uses (\texttt{SIGHASH\_ANYONECANPAY | SIGHASH\_SINGLE}) to sign the single prevout and a single output related to the input where it is verified. The program can use $S_2$ to obtain the authenticated fields for the prevout hash and index, and then check with $S_1$ that \texttt{sha\_prevouts} matches the hash of the single prevout obtained from $S_2$, and no other input is involved. Finally, if any signature uses an invalid sighash flag, this can be easily checked by the program.
\end{itemize}

\subsubsection{Invalid V}
\begin{itemize}
    \item Program inputs: $S_1$, $C'$, and $V$.
    \item Goal: Bob wants to prove that the program input hash V signed with $O^A_V$ does not match the data signed with Schnorr in signature $S_1$.
    \item Checks: First, the BitVMX program must verify that the signature $S_1$ is valid against C' provided by Bob. If not, then Bob loses. The program can then perform one of two equivalent checks. This first requires Bob to provide X and L. The program can check that X is associated with a Merkle Tree which has the script hash L on one of the leaves. The value L must be provided by Bob together with the OT-signed Merkle path of the leaf containing L. The program must verify this path. From the previous fraud check we know that \texttt{sha\_outputs} is correct, and the output containing X is well formed. Bob wins if this L does not match V. A simpler approach is to extract L from \texttt{tapleaf\_hash} in the script path spend extension which exists in C'. We use this method.
\end{itemize}

\subsubsection{Invalid W}
\begin{itemize}
    \item Program inputs: $W$, $S_1$, $C^A$. 
    \item Bob wants to prove that the value W given in $C^A$ does not correspond to a valid signature for transaction $P^B_R$. 
    \item Checks: First, the program checks $S_1$ against the transaction $C^A$ provided. If not, the fraud proof fails. 
    The template of the transaction $P^B_R$ is hardcoded in the fraud proof verification script, including the required public keys and its second input prevout hash and index. However, since the first input of transaction $P^B_R$ depends on the ID of transaction $C^A$, Bob must bring $C^A$ to prove the mismatch in ID for its first input. Even if we need to provide $C^A$ (without the witness) to compute its transaction ID, we already know that $C^A$ is small ($\sim$200 vbytes), since $C^A$ is well formed (if not, then the first fraud is used). Once the program builds the transaction using the the template and the transaction ID of $C^A$, it can check the signature $W$. Bob wins if the signature fails to pass the verification.

    If we want to protect the protocol against transaction replacement cycling attacks, we can add a fourth fraud proof. If not, then the sole existence of R proves that the PI has been revealed. We can either penalize Alice for actually performing an attack or simply penalize Alice for using a transaction template that may enable a future attack.
\end{itemize}

\subsubsection{Invalid $R^A$}
\begin{itemize}
    \item Program inputs (to prevent grinding by restricting I/O): $R^{A'}$, $Y$, $E_R$. Here $R^{A'}$ is the signed message for $R^A$, $Y$ is Alice's signature of this structure, and $E_R$ contains all remaining fields in the signed message that are required to verify the signature $Y$ of $R^{A}$.
    \item Program inputs (to prove that a grinding attack was performed): $G_1$, $G_2$, $Y_1$, $Y_2$. 
    \item Goal: Bob wants to prove that the transaction $R^A$ has no additional inputs or outputs that facilitate its grinding to perform transaction replacement cycling attacks. If we cannot restrict the outputs of transaction $R^A$ (because we need to allow CPFP or the inputs of $R^A$ to pay for transaction fees), then we may at least let Bob prove that Alice performed a grinding attack.
    \item Checks: To prevent grinding, we assume that the template for $R^A$ is fixed and hardcoded. The program checks that the inputs and outputs are exactly the ones in the template (i.e. one input and one dummy unspendable output). The program must also check that the signature Y uses the \texttt{SIGHASH\_ALL} flag. 

    If our protocol needs to allow $R^A$ to have additional inputs/outputs, then Bob may want to prove that Alice has performed a grinding attack. When Bob detects two different transactions $R_1$ and $R_2$, consuming the same output of $D^A$ but with two different signatures $Y_1$ and $Y_2$, Bob can penalize Alice. Let $G_1=\text{SHA-256}(R'_1)$ and $G_2 = \text{SHA-256}(R'_2)$. All $Y_1$, $Y_2$, $G_1$ and $G_2$ are program inputs. The program must first check that both signatures are valid with a raw Schnorr verification (no pre-hash), with respect to the hashes $G_1$ and $G_2$, and fraud is proven if the $G_1$ is not equal to $G_2$.
\end{itemize}

\subsection{Security Analysis}
Due to the complexities of Bitcoin consensus rules regarding P2SH and P2TP and the reliance of the security of this protocol in consensus rule details, a formal model covering the whole protocol is out of the scope of this work, and abstracting out the details to prove soundness would provide a false sense of security. Therefore we present a list of all attack vectors we have identified and how the protocol resists them.
\begin{enumerate}
    \item \texttt{C\_MIS}: Alice doesn't publish transaction C. In this case the handle times-out and Bob can issue a transaction PC, which stops the protocol and prevents Alice from taking control of the funds.

    \item \texttt{C\_INV\_OUT}: Alice publishes a transaction C with more than one output. Bob can punish Alice with the \textit{invalid transaction CA} challenge.

    \item \texttt{C\_INV\_SH}: Alice publishes a transaction C and signs $S_1$ or $S_2$ with invalid sighash flags. Bob can punish Alice with the \textit{invalid transaction CA} challenge.

    \item \texttt{C\_INV\_TT}: Alice publishes a transaction C with an output that does not match the expected taptree. Bob can punish Alice with the \textit{invalid transaction CA} challenge.

    \item \texttt{C\_INV\_IK}: Alice publishes a transaction $C^A$ with an output using a taproot internal key that is spendable with a Schnorr signature instead of a taptree. Bob can punish Alice with the \textit{invalid transaction CA} challenge.

    \item \texttt{C\_INV\_INP}: Alice publishes a transaction $C^A$ with more than one input. Bob can punish Alice with the \textit{invalid transaction CA} challenge. 

    \item \texttt{C\_INV\_ANNEX}: Alice publishes a transaction $C^A$ with a \textit{sha\_annex}. Bob can punish Alice with the \textit{invalid transaction CA} challenge. 

    \item \texttt{R\_MIS}: Alice does not publish a transaction $R^A$. In this case, the address X has a taptree leaf that times-out, and Bob can issue a transaction PR, which stops the protocol and prevents Alice from taking control of the funds.

    \item \texttt{R\_INV}: Alice publishes a transaction $R^A$ that has more inputs, more outputs, or other abnormal features. This could lead to transaction-replacement cycling attacks. If we want to protect ourselves against these attack vectors, we can do so by implementing the \textit{invalid RA} challenge.

    \item \texttt{R\_INV\_SCRIPT}: Alice publishes a script in $R^A$ that does not contain the User Input in a block surrounded by \texttt{OP\_IF}/\texttt{OP\_ENDIF} as expected or she doesn't use a single \texttt{OP\_PUSH} in between, or the script contains additional unexpected opcodes or data. The program running in the first instance of BitVMX will extract the User input from the Program Input and will detect any anomaly and halt execution. 

    \item \texttt{R\_INV\_V}: Alice publishes a program input using transaction envelopes, but the input does not match the hash digest V. Bob can punish Alice with the \textit{invalid $V$} challenge.

    \item \texttt{C\_FR}: Alice tries to prevent Bob from publishing transaction $P^B_C$ by publishing transaction $C^A$, after the timelock has expired (front-run attack). In this case, Bob can engage in trying to win over Alice using RBF, or he can simply move forward with the protocol and accept the delay, since the delay does not cause any financial loss to Bob.

    \item \texttt{R\_FR}: Alice tries to prevent Bob from publishing transaction $P^B_R$ by publishing transaction $R^A$, after the timelock has expired. Same as in the previous attack, Bob simply continues with the protocol.
\end{enumerate}

This completes the identified attack vectors and shows how the protocol is secured against them.

\section{Summary}
In this paper, we have presented a new method to sign BitVMX program inputs with ECDSA or Schnorr signatures, instead of using an OTS scheme. We achieved a $1:1$ data expansion factor, making our scheme much more efficient than the current method using the Winternitz scheme, which has a data expansion factor of $1:200$. This improvement allows BitVMX to verify uncompressed SPV proofs or longer computation integrity proofs, such as STARKs. To protect from malformed or fraudulent data publications we use a secondary BitVMX instance that verifies the correctness of signatures against the sequential hash of the program input data, then use the Winternitz signature of the sequential hash to check the data inside the BitVMX CPU. We add a SHA-256 hasher to the BitVMX CPU to hash the program input and check against the signed hash digest, together with a new trace partition method to identify an incorrect state transition while hashing the input data. Our more advanced scheme based on transaction envelopes uses standard Bitcoin transactions and has minimal overhead, while the changes in the CPU require a low amount of additional transaction space, and they do not change the dispute protocol worst case.

\end{document}